\documentclass{epl}
\usepackage{amsmath,amssymb}

\title{Thermal breakdown of coherent backscattering: a case study of
quantum duality}
\shorttitle{Thermal breakdown of CBS}

\author{Christian Wickles  \and Cord M\"uller}

\institute{                    
  Physikalisches Institut, Universit\"at Bayreuth, 95440 Bayreuth, Germany
}


\newcommand {\ein}      { \hat{\vect{\epsilon}}_\ab{in} }

\newcommand {\win}      { \omega_\ab{in} }
\newcommand {\kin}      { {\vect k_\ab{in}} }
\newcommand {\hatkin}   { \hat{\vect{k}}_\ab{in} }

\newcommand {\eout}     { \hat{\vect{\epsilon}}_\ab{out} }

\newcommand {\kout}     { {\vect k_\ab{out}} }

\newcommand {\kprime}   { {\vect k^\prime} }

\newcommand {\who}      { {\omega_\ab{ho}} }
\newcommand {\lho}      { {\lambda_\ab{ho}} }
\newcommand {\wR}       { {\omega_\ab{rec}} }

\newcommand {\vRMS}     { {v_\ab{rms}} }

\newcommand {\Rhat}   { \hat{\vect {R}_0} }

\newcommand {\Rtwo}   { {\vect R}_2 }
\newcommand {\Rone}   { {\vect R}_1 }
\newcommand {\Rtwone} { {\vect R}_{21}}

\newcommand {\tr} {\mathrm{tr}}
\newcommand {\rhoInitial} { {\rho^{(\ab{i})} }}

\newcommand{\ket}[1]{\lvert#1\rangle}
\newcommand{\bra}[1]{\langle #1\rvert}
\newcommand{\abs}[1]{\lvert#1\rvert}
\newcommand{\bbeta}{\bar{\beta}}

\begin{document}

\maketitle

\begin{abstract}
We investigate coherent backscattering of light by two harmonically
trapped atoms in the light of quantitative quantum
duality. Including recoil and Doppler shift close to an optical
resonance, we calculate the interference
visibility as well as the amount of which-path information,
 both for zero and finite temperature. 

\end{abstract}

\section{Introduction} 

A wealth of information about the motion of microscopic particles 
can be gathered by scattering a well-controlled probe, typically
light in the case of atoms with optical resonances. 
Young's historic double slit
experiment demonstrated that the probe light wave shows 
interference if it can propagate along more than a single path. 
Interference fringes therefore can give sensitive information about
the scatterers, as was shown by Eichmann and coworkers 
\cite{Eichmann} by realizing Young's double slits with two
trapped atoms. On the other hand, the motional
state of the probed particle gives information about the probe field, as
recently discussed by Eschner \cite{Eschner03}.
As a commonly accepted rule, one can have either full interference contrast or
full which-way-information at the same
time. Between these two extremes, there are interesting intermediate
situations that can be quantified using general 
quantum duality relations derived by Englert
\cite{Englert} that have been used for instance in 
atom interferometers \cite{Duerr98}.  
In mesoscopic samples of weakly
disordered clouds of cold atoms, coherent multiple scattering of light 
leads to enhanced backscattering \cite{CBS,Labeyrie00}. 

In this Letter, we investigate coherent backscattering (CBS) of light by two atoms that are
trapped in harmonic oscillators. This model system can describe how 
atomic motion destroys multiple scattering interference via recoil and
Doppler effects. We calculate the CBS interference visibility both at zero and finite
temperature in shallow traps that allow to treat the limiting case of freely moving atoms. 
The CBS double scattering geometry realizes a two-way interferometer
 where the two mutually exclusive alternatives
are the order in which the photon visits both atoms. 
Which-way information is then present if one knows which of both atoms has
been visited first. We calculate the which-way 
distinguishability proposed by Englert 
and show how it is physically encoded. 


\section{The setting} 

\begin{figure}
\begin{center}
\includegraphics[width=50mm]{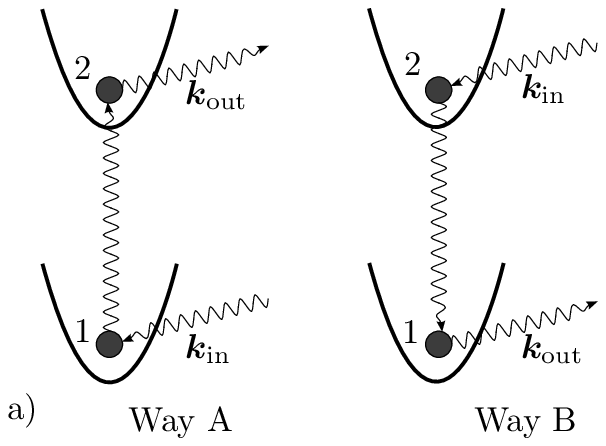}
\hspace{3cm}
\includegraphics[width=40mm]{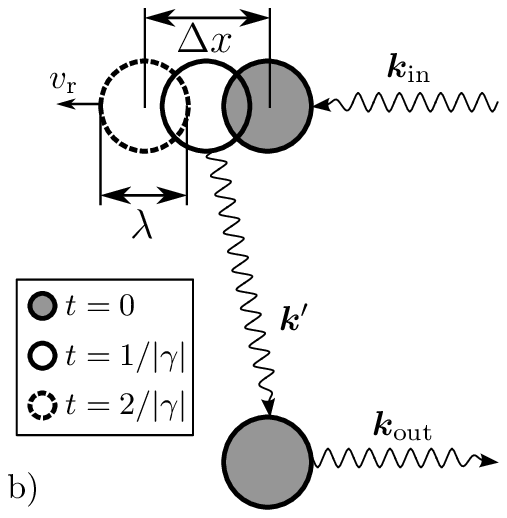}
\caption{(a) Two 
harmonically trapped atoms 
backscatter a probe photon along way A or B.
(b) Recoil mechanism: which-way information can be extracted if the wave-packet
displacement $\Delta x=v_\ab{rec}\Delta t$ of the first scatterer 
during the total scattering time $\Delta t$ is larger than the initial position
uncertainty $\lambda$. 
}
\label{twoways.fig}
\label{atomicOverlap.fig}
\end{center}
\end{figure}

Consider two atoms trapped in identical 
harmonic oscillators at fixed positions well separated from each other by many
wavelengths of the probe light. 
A single incident photon with wave vector $\kin$ and polarization
$\ein$ is then scattered by the atoms and detected in the
backscattering direction $\kout=-\kin$ with polarization analysis in
the helicity-preserving channel $\eout=\ein^\ast$. 
 A photon scattered in the backscattering direction by a single atom with a non-degenerate ground state 
 has the same polarization, but
opposite helicity, and does not contribute to the detected intensity. 
Since the atoms are far from each other, the probability for repeated scattering is very small. 
The detector then receives a photon that has been
scattered exactly once by each atom, either along way A or along way
B, see fig.~\ref{twoways.fig}. 
The total amplitude is the coherent superposition of both amplitudes. 
Such an interference of counter-propagating multiple scattering amplitudes
in the backscattering direction is
known as coherent backscattering (CBS) 
\cite{CBS}. 

Without interaction, the free electromagnetic field and the internal and external (motional)
atomic degrees of freedom are described by  the Hamiltonian $H_0 =  H_\ab{ph} + H_\ab{int} +
H_\ab{ext}$ or 
\begin{equation}
H_0 =  \sum_\mu \hbar \omega_\mu a_\mu^\dagger a_\mu 
+ \hbar \omega_0 \left (\mathrm{P}_{\ab{e}1} + \mathrm{P}_{\ab{e}2} \right )
+ \hbar \who ( N_1 + N_2 ).
\label{freeHamiltonian}
\end{equation}
The first term constitutes the standard Hamiltonian for free photons
\cite{Loudon}. 
$\mathrm{P}_{\ab{e}i} =\sum_{m_i}\ket{1m_i}\bra{1m_i}$ is the projector onto the degenerate
multiplet $J_\ab{e} = 1$ of excited states 
with transition frequency $\omega_0$ from the non-degenerate
ground state $J_\ab{g} = 0$ of atoms $i=1,2$.  
The atomic motion is described by the number
operators $N_i= N_{ix} + N_{iy} + N_{iz} = \vect a_i^\dagger \vect
a_i$ of excitations in the three-dimensional isotropic harmonic
oscillators with frequency $\who$. 

The dipole interaction 
$V=-{\vect D}_1\cdot{\vect E}({\vect R}_1) -{\vect D}_2\cdot{\vect E}({\vect R}_2)$
couples the photon field both to the electronic states and the motional degrees of
freedom. $ \vect D_i $ is the electronic dipole transition operator  
for atom $i$. The electric field operator $\vect{E} (\vect R)$ 
is evaluated at the atomic center-of-mass position  
$\vect R_i = \vect R_i^{(0)} +
\vect u_i$. The displacement operator $ \vect u_i = \lho ( \vect
a_i^\dagger + \vect a_i)$ from the trap's origin  
$\vect R_i^{(0)}$  measures distance in units of the oscillator length 
$\lho =  \sqrt{\smash[b]{\hbar/(2m\who)}}$.  
The vector joining the two atoms will be denoted $\Rtwone=\vect
R_2-\vect R_1$, with a distance $\vect R_0=\vect
R_2^{(0)}-\vect R_1^{(0)}$ between the traps such that $k_\text{in} R_0\gg1$.


\section{Transition operator for photon double scattering}

Since the atoms are well separated, the total double scattering
process is described by the product of two individual scattering
processes with free propagation inbetween.
In the far-field photon propagator $ {e^{i k^\prime
\abs{\Rtwone}}}/\abs{\Rtwone}$, 
we expand the absolute distance to linear order in the small relative
displacement: 
$\abs{\Rtwone} \approx R_0 + \Rhat \left( {\vect u_2} - {\vect u_1}
\right)$. 
In the exponential, the zeroth-order contribution $R_0$ drops out from
all interference quantities, whereas the linear term
generates a phase difference between the amplitudes of
$A$ and $B$ and must be kept. 
To this leading order, the denominator can be taken constant.  

The transition operator 
\cite{CCT} 
for way A takes a useful form in time representation and interaction picture
where $\vect R_j(t) = e^{iH_0 t/\hbar} \vect R_j e^{-iH_0 t/\hbar}$. Up to
irrelevant prefactors, 
\begin{equation}
T_\ab{A}  = 
\int_0^\infty \!\!\!\int_0^\infty  
e^{i\gamma\left(s+t\right)} 
e^{-i\kout\Rtwo(0)} e^{i \kprime \Rtwo(-s)}
e^{-i \kprime \Rone(-s)}
e^{i \kin \Rone(-s-t)}
\,\upd t\,\upd s  .
\label{transitionOperatorA}
\end{equation}
Read from right to left, it describes how the photon is scattered
first by atom 1, then propagates with $\kprime \equiv k_\ab{in}
\Rhat$, and is finally scattered by atom 2. 
The transition operator $T_\ab{B} $ of the reverse way B is 
obtained from $T_\ab{A}$ by the substitution 
$ \Rone\leftrightarrow\Rtwo,\kprime\rightarrow-\kprime$.
The complex detuning $\gamma = \win-\omega_0 +i\Gamma/2$ of
the probe frequency $\win$ from the transition frequency 
includes the spontaneous decay rate or inverse
lifetime $\Gamma$.
Retardation times 
of order $R_0/c$ have been neglected in the time
arguments of the operators. Indeed, 
the photon scattering by two atoms defines two distinct time scales for the atomic motion:
first, the total inverse detuning $|\gamma|^{-1}$ which is simply  
 $2/\Gamma$ at resonance. 
Second,  the free propagation time $R_0/c$ from one atom to the other. 
For resonant atomic scatterers with $1/\Gamma$ of order $10^{7}\,$s or larger,
and typical distances $R_0$ of 1$\,$mm or less,
one has $R_0/c\ll 1/\Gamma$ such that the scattering is
resonance-dominated, and the free propagation
time can be safely neglected.  
In all of the following, we have in mind the limit of quasi-free atoms
and therefore consider the case of shallow traps $\who \ll \Gamma$ in
which the oscillation period of an atom is much larger than the
time $\Gamma^{-1}$ it takes to scatter a photon.

\section{Quantitative quantum duality}

The incident photon can choose between two \textit{a priori} indistinguishable 
paths, way $\ab{A}$ or $\ab{B}$. 
Following Englert's fashionable choice \cite{Englert},
we call this binary degree of freedom a qubit. Let 
$|\ab{A}\rangle$ and $|\ab{B}\rangle$ denote the choice of way
$\ab{A}$ and $\ab{B}$. 
Scattering entangles the qubit with the motional
degrees of freedom. Consequently, the atomic oscillator states 
can serve as a  which-path detector. 

Prior to scattering, the total initial state of qubit 
and detector is 
$\rho_\ab{tot}^{(\ab{i})} = \rho_\ab{AB}^{(\ab{i})} \otimes
\rhoInitial$. 
The qubit is in the pure state $\rho_\ab{AB}^{(\ab{i})} = \frac 1 2 \left( |\ab{A}\rangle + |\ab{B}\rangle \right) 
\left( \langle \ab{A}| + \langle \ab{B}| \right)$, the symmetric superposition of equally probable ways. 
An external cooling laser field serves as a thermal bath 
for the trapped atoms. The detector is thus in a
thermal state $\rhoInitial = e^{-\beta H_\ab{ext}}/Z$ at inverse temperature $\beta=1/k_\ab{B}T$ with
partition function $Z = \tr\{e^{-\beta H_\ab{ext}} \}$.  
The total final state is then obtained by applying the transition operators associated with way
$\ab{A}$ and $\ab{B}$:  
\begin{equation}
\rho_\ab{tot}^{(\ab{f})} = \frac{T\rho_\ab{tot}^{(\ab{i})} T^\dagger}{I_\ab{A} + I_\ab{B}} ,\qquad 
T = |A\rangle \langle A| \otimes T_\ab{A} + |B\rangle \langle B|
\otimes T_\ab{B}  .
\end{equation}
In general, $T$ is not a unitary operator since
it describes only the scattering amplitude around the backscattering
direction. The factor
$I_\ab{A}+I_\ab{B} \equiv \tr \{ T_\ab{A} \rho^{(\ab{i})}
T_\ab{A}^\dagger \} +
\tr \{ T_\ab{B} \rho^{(\ab{i})} T_\ab{B}^\dagger \}$
guarantees, however,  that $\rho_\ab{tot}^{(\ab{f})}$  is properly normalized.

Adapting Englert's general definitions \cite{Englert}, we
can express the
{\it visibility} ${\cal V}$ and the {\it distinguishability} ${\cal D}$ 
obeying the fundamental {\it duality relation} ${\cal V}^2 +
{\cal D}^2 \leq 1 $ as follows:
\begin{equation}
{\cal V} = 2 \frac {\left| \tr  \{ T_\ab{B} \rhoInitial
T_\ab{A}^\dagger \} \right| } {I_\ab{A} + I_\ab{B}}
, \qquad 
{\cal D} = \frac {\tr \left| T_\ab{A} \rhoInitial T_\ab{A}^\dagger -
T_\ab{B} \rhoInitial T_\ab{B}^\dagger \right|  } {I_\ab{A} +
I_\ab{B}}. 
\label{D_and_V}
\end{equation}
The visibility is the ratio of interference contribution and background 
intensity $I_\ab{A} + I_\ab{B}$ and therefore
equal to the CBS contrast.  
The distinguishability ${\cal D}$ describes the
maximum which-way information available in principle, i.e., that
can be extracted by the additional measurement of an optimal
detector observable; 
$\tr \left| X \right| = \tr \{ \sqrt{X^\dagger X}\}$
denotes the trace-class norm of the operator $X$. 
Another interesting quantity is  
the {\it predictability}  
\begin{equation}
{\cal P} =  \frac { \left| \tr \left\{ T_\ab{A} \rhoInitial T_\ab{A}^\dagger -
T_\ab{B} \rhoInitial T_\ab{B}^\dagger \right\} \right| } {I_\ab{A} +
I_\ab{B}} =  \frac {|I_\ab{A}- I_\ab{B}|} {I_\ab{A} + I_\ab{B}} .
\label{P}
\end{equation}
The predictability measures the amount of which-way information
available {\it a priori}, as for instance in unbalanced
interferometers like Young's double slits with different widths. For
balanced interferometers $I_\ab{A}=I_\ab{B}$, the predictability
vanishes, $\mathcal{P}=0$. 

With the help of these quantities, we can quantify the breakdown of coherent
photon backscattering by mobile atoms, both at zero and finite temperature.  
The distinguishability $\cal D$ is difficult to evaluate in our case
because it is defined via the trace-class norm of operators on the 
infinite-dimensional Hilbert space of harmonic oscillators. On the
contrary, visibility ${\cal V}$ and predictability ${\cal
P}$ involve a thermal harmonic average of products like $T_\ab{A}^\dagger
T_\ab{B}$ or $T_\ab{A}^\dagger T_\ab{A}$. Since the transition operator
(\ref{transitionOperatorA}) 
contains only exponentials linear in the displacement, 
taking the trace amounts to Gaussian integration.

\section{Zero temperature}

At zero temperature, the atoms are initially in their respective
harmonic oscillator ground states. In other words, the detectors are
prepared in pure states. Setting
$\rhoInitial=|\Psi\rangle\langle\Psi|$ in (\ref{D_and_V}) permits to show that in
this case the duality
relation is always saturated: ${\cal D}^2 = 1 - {\cal V}^2$. 
Thus, we can use the visibility ${\cal V}$, much easier to
calculate than the distinguishability ${\cal D}$,  in order to 
understand how much which-way information is present and how it is encoded. 
Carrying out the thermal average, we find in the limit $T\to 0$
\begin{eqnarray}
{\cal D}^2  &=& {\varrho}^2(\win) 
             + 2 \zeta^2(\win) = 1-{\cal V}^2 ,
\label{DistZero}\\
{\cal P}^2 &=&  {\varrho}^2(\win),
\label{PredZero}
\end{eqnarray}
neglecting higher-order terms $O(\zeta^4,\chi^3,\zeta^2\chi)$ in
the two relevant small parameters defined as follows:    
The influence of atomic recoil without harmonic trapping 
is encoded in the factor 
\begin{equation}
\varrho(\win)
= 4 (\Rhat\cdot\hatkin) \frac \delta {\abs\gamma} \chi,
\quad\quad \chi\equiv \frac \wR {\abs\gamma}, 
\label{varrho}
\end{equation}
with  $\hbar\wR = \hbar^2 
k_\ab{in}^2 / (2 m)$ the recoil energy, $\delta = \win-\omega_0$ 
the probe detuning from the atomic resonance, 
and  $\gamma= \delta+ i\Gamma/2$ as before.  
The harmonic trap enters through the parameter   
\begin{equation}
\zeta^2(\win)\equiv \frac{4\wR \who}{|\gamma|^2} .
\label{zeta}
\end{equation}
It is worthwile to discuss the physical significance of both
parameters.  

Let us first consider a very shallow trap  $\who\ll \wR$ where the free
recoil effect described by $\varrho(\win)$ dominates. At least to
order $\chi^2$, we find ${\cal P}^2 + {\cal V}^2=1$. Together with the
general property ${\cal P} \le {\cal D}$ 
(read: the \textit{a priori} information cannot be 
larger than the total available information)    
this implies that all which-way information is 
actually  available {\it a priori}:
${\cal D}={\cal P}$. 
Just as in the case of other asymmetric interferometers, this predictability is due to
unbalanced scattering amplitudes.
At perpendicular scattering $\Rhat\cdot\hatkin=0$, the situation is completely
symmetric, and both paths are equally probable. 
But for the extreme case $\Rhat\cdot\hatkin=\pm 1$ of atoms in line with the probe, 
the atom in front scatters either the
incident probe photon at the laser frequency or the already scattered photon on its way out
again,  now at laser frequency
minus twice the recoil. A different frequency generally implies a 
different scattering cross section such that the two paths have
different probabilities.
 For a given
position configuration, this information
is known \textit{a priori} without the necessity to perform any measurement. 
More quantitatively, the relative
change in the resonant cross section $\sigma(\delta)=\sigma_0
[1+(2\delta/\Gamma)^2 ]^{-1}$ 
under a small frequency change $\Delta \omega\ll \Gamma$ then is
$|\Delta\sigma/\sigma| =  \Delta \omega\,  2\delta /
|\gamma|^2$, 
see fig.~\ref{visibility.fig}(a). 
For the actual frequency change $\Delta\omega = 2 (\Rhat\cdot\hatkin)\wR$, 
one finds exactly $|\Delta\sigma/\sigma| = \varrho(\win)$.  
To this order, the predictability vanishes at exact detuning, since a small frequency 
change on the flat top of the 
resonance Lorentzian has no effect.

Let us now interpret the influence of harmonic trapping at zero
temperature. The predictability (\ref{PredZero}) contains no
contribution in $\zeta^2(\win)$, which indicates
that this parameter encodes which-path information
that may only be revealed \textit{a posteriori} by an
appropriate measurement on the detector.
In a temporal picture (see fig.~\ref{atomicOverlap.fig}(b)), 
one can determine which way the photon has taken if one
can measure which atom has scattered the photon first. This is only
possible if the initial position uncertainty $\lambda$ is smaller than the
distance $\Delta x$ travelled by the first scattering
atom. 
More quantitatively, 
the first scatterer takes up the recoil in the direction of $\hatkin$ and travels the distance  
$\Delta x = v_\ab{rec}\Delta t = \sqrt{\smash[b]{2\hbar\wR / m}}\Delta t $ during the whole time
$\Delta t= 2|\gamma|^{-1}$ until the emission of the final photon by the
second atom. 
\footnote{$\Delta t $ is simply twice the total contribution per
atom that can be justified by a stationary-phase argument if both
the Wigner time delay of the phase and the amplitude change of the resonant
scattering function $t(\delta)=t_0/(\delta+i\Gamma/2)$ are taken into
account.}
At that moment, the second atom receives the same recoil in the direction  $\hatkin$,
such that the final momentum states for both ways become 
indistinguishable (note that the recoil in the direction joining the
atoms is exchanged instantaneously, because we can neglect the
propagation time). 
The zero-temperature position uncertainty 
is the oscillator length $\lho$, 
which finally shows that which-path information is indeed present on the 
scale $\Delta x/\lho = 2 \zeta(\win)$.

\section{Finite temperatures}

\begin{figure}
\begin{center}
\includegraphics[width=55mm]{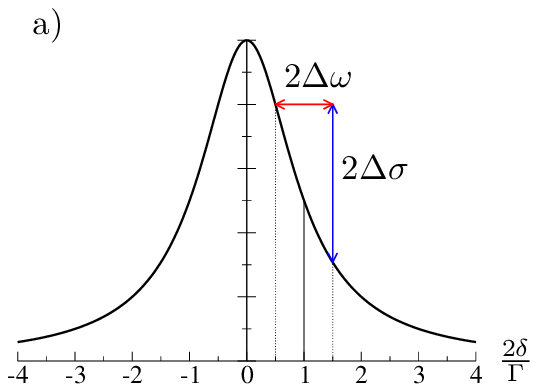}
\hfill
\includegraphics[width=75mm]{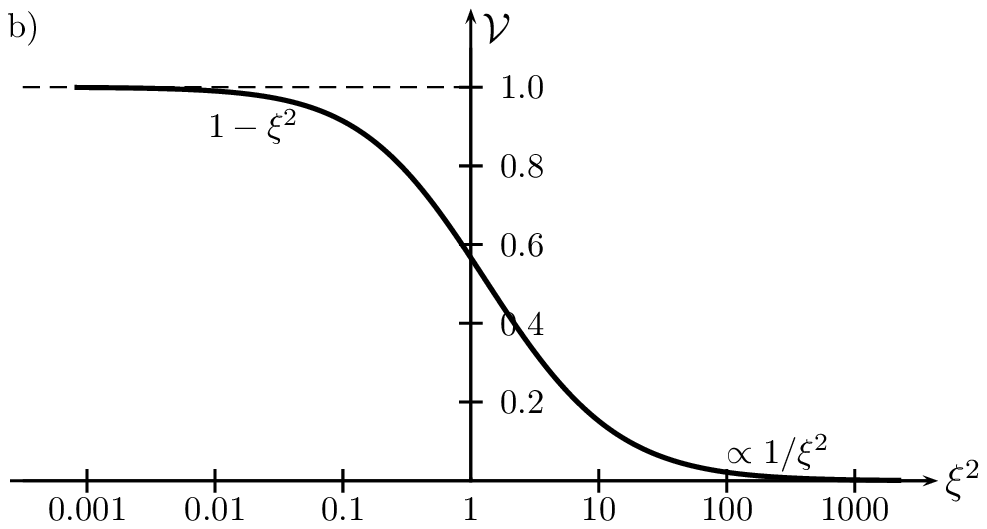}
\caption{(a) Frequency-shift mechanism: a small frequency change
$\Delta\omega$ on the steep slope of a resonance curve implies a large
scattering cross section difference $\Delta\sigma$ that can provide
which-path information. 
(b)  CBS visibility (\ref{D_and_V}) as
function of the thermal
Doppler shift (\ref{xiclassical}) at resonance $\delta=0$. 
}
\label{visibility.fig}
\end{center}
\end{figure} 

If the atoms are coupled to a thermal bath, the detector is not
prepared in a pure state, and one expects that the duality
relation no longer saturates. We thus have to calculate visibility
and distinguishability separately. 

After the Gaussian thermal average, the visibility 
is given in terms of a product 
of amplitudes like (\ref{transitionOperatorA}) for way A and
B. It can be evaluated numerically
for all values of the detuning $\delta$ 
and of the thermal trap parameter 
$\xi^2(T,\win) \equiv \coth(\frac 1 2 \beta\hbar\who) 
\zeta^2(\win).$
This parameter corresponds to the 
zero-temperature trap parameter (\ref{zeta}) at effective oscillator
excitation $[2n(T)+1]\who$, with $n(T)=[e^{\beta\hbar\who}-1]^{-1}$ the 
Bose-Einstein distribution function.
For highly excited atoms $\beta\hbar\who\ll 1$ 
the oscillator frequency drops out of the definition of the thermal
trap parameter which becomes  
\begin{equation}
\xi_\ab{cl} = \frac{2 k_\ab{in} v_\ab{rms}}{|\gamma|},
\label{xiclassical} 
\end{equation}
the classical thermal Doppler shift ($v_\ab{rms}^2 = \langle {\vect v}^2\rangle$) in
units of the linewidth. 
Figure \ref{visibility.fig}(b) shows a plot of the visibility 
at exact resonance $\delta=0$ as function of $\xi_\ab{cl}^2$
on a semi-logarithmic scale.  
Here, we concentrate on the effect of harmonic
trapping at finite temperature and neglect the free recoil contribution, which is
justified if $\chi/\xi_\ab{cl}^2 \ll 1$.
The visibility decreases monotonically with temperature. 
For  $\xi_\ab{cl}^2\ll 1$, we can expand to lowest order and find
analytically (compare with eq.~(\ref{DistZero})): 
\begin{equation}
{\cal V}^2  = 1 - 2 \xi_\ab{cl}^2, 
\label{VisFinal}
\end{equation}
At this point, we recover the case of free thermal atoms. Our result
(\ref{VisFinal}) agrees with the CBS contrast calculated 
in \cite{Wilkowski03} for the low-temperature case $\xi_\ab{cl}\ll 1$ 
(if the avarage medium effect included there is disregarded). Note
that our calculation is also valid in the high-temperature regime
$\xi_\ab{cl}\gg 1$ where the visibility goes to zero as
$\mathcal{V}\sim \xi^{-2}$. 

It is too difficult to evaluate analytically the trace-class norm 
for the distinguishability ${\cal D}$ 
with the transition operators in the general form
(\ref{transitionOperatorA}). 
Therefore, we formally expand the exponentials 
in powers of the Lamb-Dicke parameter $k_\ab{in} \lho = \sqrt{\smash[b]{\wR / \who}}$.
The leading order expression for ${\cal D}$ in the 
shallow-trap limit
$\beta\hbar\who\ll 1$ becomes 
\begin{equation}
{\cal D} = \frac{2 \sqrt 2 |\delta| }
{|\gamma|^2} 
{v_\ab{rec} \beta \who}\tr \left|e^{-\beta\hbar\who {\hat a}^\dagger {\hat a}} {\hat p} \right| .
\label{expressionD_XP}
\end{equation}
Here, ${\hat p}=(\hbar/2i\lho)[\hat a-\hat a^\dagger]$ is the momentum
associated with ${\hat a}=(\vect a_1-\vect a_2)\hatkin$, the antisymmetric 
oscillator mode projected onto the probe direction $\hatkin$.
It is reasonable that only this mode should be relevant:  
symmetric motion cannot encode differential information about the
paths (and could be disposed of by transformation into a co-moving frame),
whereas the momentum along the only other available direction $\Rhat$
is exchanged instantaneously.    
To linear order in 
$\bbeta=\beta\hbar\who$, the  positive operator $V=e^{-\bbeta {\hat a}^\dagger {\hat a}}$  
can be moved outside the absolute value 
(using $\left| V \hat p \right| \approx \sqrt{(V^\dagger V)(\hat p^\dagger \hat p)}
\approx |V| |\hat p|$ since $\hat p$ and $V$ commute to zeroth order in $\bbeta$). 
This makes the 
distinguishability proportional to the thermal expectation value of the 
momentum modulus, which is easy to evaluate for a
Maxwell-Boltzmann distribution: 
\begin{equation}
{\cal D} = \sqrt{2} \frac{|\delta|}{|\gamma|} \frac {2 v_\ab{rec}} {|\gamma|}
\frac{\langle | {\hat p} | \rangle}{\hbar} 
=  \frac{2}{\sqrt{\pi}} \frac{|\delta|}{|\gamma|}
\xi_\ab{cl}
\label{expressionD_Final}
\end{equation}
It can be read as the relative shift 
${\cal D} = \pi^{-1/2} |\Delta \sigma/\sigma|$ of the
resonant cross section under the small Doppler frequency change
$\Delta\omega = 2 k_\ab{in} v_\ab{rms}$.  
This suggests an interpretation in terms of the
frequency-shift mechanism (cf.\ fig.\ref{visibility.fig}(a)): 
By measuring the actual atomic velocities \textit{a posteriori}, one
can determine which way was more probable: if atom 1 is found to move towards a red-detuned
probe and atom 2 to move away from it, way A 
was more probable than B because atom 1 had
a larger absorption cross section---this is simply the principle of
Doppler cooling. 
Contrary to the zero-temperature case, now the predictability $\mathcal{P}$ 
is zero to the order considered because the interferometer symmetry is
reestablished by the isotropic thermal average. 
Naturally, as in the case of harmonic
trapping at zero temperature,
the distinguishability vanishes also at exact resonance
$\delta=0$ because no differential
information can be encoded on the flat top of the resonance Lorentzian.
In the high-temperature limit $\xi_\text{cl}\gg 1$, i.e., for very large frequence shifts
$\Delta\omega\gg \Gamma$ with moderate detuning, one explores the flat wings of the 
Lorentzian, and the distinguishability must then drop to zero. 

At finite temperature, the quantum duality is no longer saturated and
reads to order $\xi_\ab{cl}^2$:  
\begin{equation}
{\cal V}^2 + {\cal D}^2 = 1-2 
\left[1-\frac{2}{\pi}\frac{\delta^2}{|\gamma|^2}\right]  
\left(\frac {2 k_\ab{in} \vRMS} {|\gamma|} \right)^2
  .
\end{equation}
This  value of ${\cal D}$ assumes an optimal measurement
within just the system of center-of-mass motion. But the initial
thermal distribution requires the presence of a thermal bath 
(e.g., cooling lasers). The total system (detector plus bath) is
described by a pure state for which the quantum duality
saturates. However, the which-path information could only be retrieved 
by an optimal measurement including all bath degrees
of freedom that are not under our control. 
Measurements on just the detector necessarily correspond to a
non-optimal measurement with respect to the whole system and therefore
imply a reduced  distinguishability.

\section{Summary}

We have studied the coherent backscattering effect from two trapped
atoms, an interesting case study for quantitative quantum duality in a
physically realistic setting. At zero temperature, inelastic
scattering due to the recoil effect provides \textit{a priori} 
which-path information. 
Further which-path information can be
gained by measuring which atom has scattered the photon
first, which is possible if the initial position uncertainty is
small enough. 
At finite temperature,  the atomic 
movement destroys the interference
once the average Doppler shift becomes of the order of the resonance
width. Which-path information resides in the atomic velocities via 
resonance conditions, but the duality inequality no longer saturates 
because 
an initial thermal state corresponds to a non-optimal detector
preparation.     

We hope
that these considerations may stimulate further work, both
experimental and theoretical, on the interference probing of quantum
dynamics of trapped particles.

\acknowledgments
We gratefully acknowledge a stimulating collaboration 
with D.~Delande and C.~Miniatura, who brought Englert's work to our
attention. We further wish to thank G.~Morigi and J. Eschner for 
helpful discussions and 
O. Sigwarth and R. Kuhn  for numerous remarks and a critical reading of the
manuscript. 
This project was financially supported  by the \textsc{PROCOPE} program
of the DAAD.

\end{document}